\documentclass[fdp,a4paper,fleqn%
]{w-art}
\usepackage{times,cite,w-thm,w-greek}
\theoremstyle{plain}

\theoremstyle{definition}

\usepackage[]{graphicx}

\usepackage{amssymb,amsmath,amsfonts,amsthm,amstext,amscd,array}
\usepackage{tikz}
\usepackage{mathrsfs}
\usepackage{bbm}
\usepackage{bm}
\usepackage[arrow,matrix,curve]{xy}
\usepackage{bbding}
\usepackage{wasysym}

\def\tri{\triangleright}

\def\ii{{\,{\rm i}\,}}
\def\dd{{\rm d}}

\def\Id{{\rm id}}

\def\mfg{{\mathfrak g}}
\def\mfh{{\mathfrak h}}

\newcommand{\eq}{\begin{equation}}
\newcommand{\eqend}{\end{equation}}
\newcommand{\eqa}{\begin{eqnarray}}
\newcommand{\nonueqa}{\begin{eqnarray*}}
\newcommand{\eqaend}{\end{eqnarray}}
\newcommand{\nonueqaend}{\end{eqnarray*}}

\newcommand{\bma}[1]{\begin{array}{#1}}
\newcommand{\ema}{\end{array}}
\newcommand{\bc}{\begin{center}}
\newcommand{\ec}{\end{center}}

\renewcommand{\theequation}{\thesection.\arabic{equation}}

\newcommand{\real}{{\mathbb R}} 

\def\ot{{\, \otimes\, }}

\newif\ifold             \oldtrue

\def\be{\begin{equation}}
\def\ee{\end{equation}}
\def\bea{\begin{eqnarray}}
\def\eea{\end{eqnarray}}
\def\bd{\begin{displaymath}}
\def\ed{\end{displaymath}}

\def\s{\sigma}

\newcommand{\beq}{\begin{eqnarray}}
\newcommand{\eeq}{\end{eqnarray}}

\makeatletter
\newdimen\normalarrayskip              
\newdimen\minarrayskip                 
\normalarrayskip\baselineskip
\minarrayskip\jot
\newif\ifold             \oldtrue            
\def\arraymode{\ifold\relax\else\displaystyle\fi} 
\def\@arrayskip{\ifold\baselineskip\z@\lineskip\z@
     \else
     \baselineskip\minarrayskip\lineskip2\minarrayskip\fi}
\def\@arrayclassz{\ifcase \@lastchclass \@acolampacol \or
\@ampacol \or \or \or \@addamp \or
   \@acolampacol \or \@firstampfalse \@acol \fi
\edef\@preamble{\@preamble
  \ifcase \@chnum
     \hfil$\relax\arraymode\@sharp$\hfil
     \or $\relax\arraymode\@sharp$\hfil
     \or \hfil$\relax\arraymode\@sharp$\fi}}
\def\@array[#1]#2{\setbox\@arstrutbox=\hbox{\vrule
     height\arraystretch \ht\strutbox
     depth\arraystretch \dp\strutbox
     width\z@}\@mkpream{#2}\edef\@preamble{\halign \noexpand\@halignto
\bgroup \tabskip\z@ \@arstrut \@preamble \tabskip\z@ \cr}%
\let\@startpbox\@@startpbox \let\@endpbox\@@endpbox
  \if #1t\vtop \else \if#1b\vbox \else \vcenter \fi\fi
  \bgroup \let\par\relax
  \let\@sharp##\let\protect\relax
  \@arrayskip\@preamble}
\makeatother

\newcommand{\wg}{\wedge}

\newcommand{\p}{\partial}
\newcommand{\mc}{\mathcal}

\newcommand{\trm}{\textrm}
\newcommand{\tbf}{\textbf}

\newcommand{\ka}{\kappa}

\def\tP{{\tilde P}}

\theoremstyle{definition}

\usepackage{hyperref}
\hypersetup{colorlinks,%
citecolor=black,%
filecolor=black,%
linkcolor=black,%
urlcolor=black,%
unicode,pdftitle={Nonassociative Field Theory on Non-Geometric Spaces},pdfauthor={Dionysios Mylonas and Richard Szabo}}

\begin{document}




\begin{flushright}
EMPG--14--9
\end{flushright}

\title[Nonassociative Field Theory on Non-Geometric Spaces]{Nonassociative Field Theory on Non-Geometric Spaces}


\author[Mylonas]{Dionysios~Mylonas}
\address{Department of Mathematics, Heriot--Watt University, Edinburgh, United Kingdom}
\author[Szabo]{Richard~J.~Szabo}
\begin{abstract}
We describe quasi-Hopf twist deformations of flat closed string
compactifications with non-geometric $R$-flux using a suitable cochain twist, and construct
nonassociative deformations of fields and differential
calculus. We report on our new findings in using this formalism to construct perturbative
nonassociative field theories on these backgrounds. We describe the
modifications to the usual classification of Feynman diagrams into
planar and non-planar graphs. The example of $\varphi^4$ theory is
studied in detail and the one-loop contributions to the two-point function are calculated.\\
  
\noindent \emph{Based on talk given by D.M. at the Workshop on Noncommutative Field Theory and Gravity, Corfu, Greece, September 8--15, 2013.}
\end{abstract}
\maketitle                   






\renewcommand{\theequation}{\arabic{equation}}

\vspace{-0.5cm}

Non-geometric backgrounds arise as consistent string vacua in $p$-form
flux compactifications via T-duality transformations. Consider the
standard example of closed strings propagating in a three-torus
endowed with non-vanishing constant 3-form flux $H=\dd B$. Employing
T-duality along all three directions takes $H$ to its T-dual 3-vector
flux $R$ and results in a purely non-geometric background where
transition functions between patches cannot even be defined
locally~\cite{Lust:2010iy}. This ``$R$-space" exhibits an intriguing
nonassociative deformation of geometry which is consistent with the original nonassociative deformations of spacetime discovered in~\cite{Blumenhagen:2010hj}, where standard conformal field theory approaches were used to study closed strings propagating in a constant $H$-flux background; we refer to the lecture notes~\cite{Blumenhagen:2014sba} of these proceedings for further details of these approaches and for a more exhaustive list of references.

A geometrization of $R$-space is provided by the phase space of its
T-dual background. This geometry is induced by regarding the
fundamental degrees of freedom in the non-geometric background as
membranes in a Courant $\s$-model whose boundary dynamics are described by a closed string quasi-Poisson $\s$-model with target space the cotangent bundle of the original membrane spacetime, which can be quantized using Kontsevich's deformation quantization~\cite{Mylonas:2012pg}. This yields a nonassociative star product of fields on $R$-space that reproduces the nonassociative geometry discovered in~\cite{Lust:2010iy,Blumenhagen:2010hj}, and leads to Seiberg--Witten maps which untwist the nonassociative product to a family of associative noncommutative star products; we refer to the lectures of P.~Schupp from these proceedings for further details of this approach. It is also equivalent to a strict deformation quantization approach that is based on integrating a pertinent Lie 2-algebra to a Lie 2-group, which leads to the formulation of the nonassociative star products discussed by D.~L\"ust in these proceedings. 

These deformations can also be acquired by twisting the Hopf algebra
of symmetries of $R$-space to a quasi-Hopf algebra using a suitable
cochain twist~\cite{Mylonas:2013jha}; see also~\cite{Mylonas:2014aga}
for a review and for further references. The advantage of this
technique is that it is algorithmic in the sense that once a twist has
been found, it can be used to deform all geometric structures on
$R$-space. In this contribution we will briefly review the results
of~\cite{Mylonas:2013jha}, and then use them to report on our new
advances in formulating nonassociative perturbative dynamics of scalar field theories on $R$-space.

The general algebraic framework can be summarized as follows. Recall that a {quasi-Hopf
  algebra} $(H,\phi)$ is a weakening of the notion of Hopf algebra in
which the coproduct of $H$ is coassociative only up to a
3-cocycle $\phi \in H\ot H \ot H$, called the {associator}. In
particular, one can regard a Hopf algebra $H$ as a quasi-Hopf algebra
with trivial associator $\phi=1_H\ot 1_H \ot 1_H$, and then use an
invertible counital 2-cochain $F \in H\ot H$ to twist it into a
quasi-Hopf algebra $(H_F,\phi_F)$; the requirement
$\phi_F:= \partial^*F=1_H\ot 1_H \ot 1_H$ is precisely the condition
for $F$ to be a 2-cocycle (or Drinfel'd twist). This cochain twisting
yields nonassociative deformations of $H$-module algebras $A$ by
requiring that the action of the twisted Hopf algebra $H_F$ on $A$ is
covariant, i.e. that it preserves the binary product on the algebra $A$ (see e.g.~\cite{Majid:book}). The binary product is deformed in this way to a nonassociative star product $\star$ which yields a quantization of $A$. This is exactly the strategy that we follow in~\cite{Mylonas:2013jha} to obtain nonassociative deformations of the geometry of $R$-space.

For this, consider a $d$-dimensional manifold $M$ with trivial
cotangent bundle ${\cal M} := T^*M$ whose coordinates are $x^I = (x^i,
p_i)$ with $(x^i) \in M$ and $(p_i) \in (\real^d)^*$. Translations on
$\cal M$ are realised by the action of an abelian Lie algebra $\mfh =
\real^d \oplus ( \real^d )^*$ of dimension $2d$ on the algebra of
smooth complex functions $C^\infty({\cal M})$. The (left) action of the generators $P_i$ and $\tP^i$ of $\mfh$ is denoted by $\tri$ and is given on $f\in C^\infty({\cal M})$ by the vector fields $ P_i\tri f := \p_i f $ {and} $\tP^i \tri f :=  \tilde\p^i f $, where $\p_i=\frac{\p}{\p x^i}$ and $\tilde\p^i=\frac{\p}{\p p_i}$. The manifold $M$ is then endowed with a constant trivector $R=\frac {1}{6}\, R^{ijk}\, \p_i \wg \p_j \wg \p_k$ which is T-dual to a background $H$-flux. The presence of the $R$-flux enhances the symmetries of ${\cal M}$ and thus $\mfh$ is enlarged to the non-abelian nilpotent Lie algebra $\mfg$ of dimension $\frac{1}{2}\, d\, (d+3)$ generated by $P_i$, $\tP^i$ and $M_{ij} = -M_{ji}$, whose action on $f\in C^\infty({\cal M})$ is given by
\be
M_{ij}\tri f := p_i \, \p_j f -p_j \, \p_i f \ . \label{repM}
\ee
From \eqref{repM} it is seen that $M_{ij}$ generate non-local
coordinate transformations that mix positions with momenta and are
known as {Bopp shifts}; here they are the algebraic analogs of T-duality transformations, in the sense that it is precisely this mixing of coordinates and momenta that makes $R$-space non-geometric.

The pertinent Hopf algebra $H$ is constructed by equiping the universal enveloping algebra $U(\mfg)$ with the standard coalgebra structure. It can also be regarded as a quasi-Hopf algebra with trivial associator $\phi=1_H \ot 1_H \ot 1_H$. Then the unique non-abelian twist $F\in H\ot H$ that can be constructed from this data is given by
\be
F= \exp{\big[ \mbox{$-\frac{\ii\hbar}{2}$}\, \big(\mbox{$\frac 14$}\,R^{ijk}\, (P_i \ot M_{jk} + M_{jk} \ot P_i ) + P_i \ot \tilde P^i - \tilde P^i \ot P_i\big)\big]} =: F_{(1)}\otimes F_{(2)} \ . \label{mcF}
\ee
This twist element is an invertible counital 2-cochain, and thus it defines a quasi-Hopf algebra $(H_F, \phi_F)$ with twisted associator $\phi_F\in H \ot H\ot H$ given by
\be
\phi_F =\p^* F= F_{23} \, \big[(\Id_H\ot\Delta)\,
F\big]\,\phi\,\big[(\Delta\ot\Id_H )\, F^{-1}\big] \, F_{12}^{-1} = \exp\big(\mbox{$\frac{\hbar^2}{2}$}\,R^{ijk}\, P_i \ot P_j \ot P_k \big) \ , \label{associator}
\ee
where $F_{23}=1_H\ot F$ and $F_{12}^{-1}=F^{-1}\ot1_H$.

With this data at hand, any left $H$-module algebra $A$ can be deformed using the star product
\be 
a\star b = \cdot\, \big( F^{-1}\tri (a\ot b)\big)  = \big( F_{(1)}^{-1}\tri a\big) \cdot\big(F_{(2)}^{-1} \tri b \big) \label{star}
\ee
for $a,b \in A$. For the algebra of functions on ${\cal M}$, by setting $A=C^\infty({\cal M})$ with pointwise multiplication, the star product calculated from \eqref{star} is given by
\be
f \star g =\cdot\, \Big(\exp{\big[ \mbox{$\frac{\ii\hbar}{2}$}\,\big( R^{ijk}\, p_k \,  \p_i \otimes \p_j + \p_i \otimes \tilde\p^i - \tilde\p^i \otimes \p_i \big)\big]}(f \otimes g ) \Big)\ . \label{sp}
\ee
This is a {nonassociative star product} which yields a nonassociative deformation quantization of the algebra of functions on $\cal M$; nonassociativity here is quantified by the action of the associator \eqref{associator} on the triple phase space star product as
\be 
(f\star g)\star h = \phi_F\triangleright\big( f\star (g\star h) \big):= \star\big[\phi_F\triangleright \big(f\otimes (g\otimes h)\big)\big] \ ,
\ee
and it arises only along the coordinate directions $M$.
The star product \eqref{sp} was originally derived in~\cite{Mylonas:2012pg} using Kontsevich's deformation quantization; it is the simplest nonassociative variant of the Moyal product.

Consider now the exterior algebra $\Omega^\bullet({\cal M})$ of smooth
complex differential forms on ${\cal M}$. Its quantization is obtained
in the same way once a suitable $H$-module structure is
identified. The action of $H$ on $\Omega^\bullet({\cal M})$ is
determined by finding the action on 1-forms and then extending it to
all forms in $\Omega^\bullet({\cal M})$ as an algebra homomorphism using the Leibnitz rule for the exterior product. For this, we require that the exterior derivative $\dd$ is equivariant under the action of $H$ in the sense that $ \dd( h\tri\omega)= h \tri ( \dd\omega) $ for all $\omega\in \Omega^\bullet({\cal M})$ and $h\in H$. Then the action of $H$ on $\Omega^\bullet({\cal M})$ is given by the Lie derivative $\mc L_h$ along an element $h\in H$, and in particular acting on the generating $1$-forms it gives $M_{ij} \tri\dd x^k := \mc L_{M_{ij}}(\dd x^k) = \delta_j{}^k \,\dd p_i -\delta_i{}^k \, \dd p_j$, while all other generators $\dd x^I$ are invariant under the action of $H$. The deformation quantization of $\Omega^\bullet({\cal M})$ is then given by setting $A=\Omega^\bullet({\cal M})$ with the exterior product, and using \eqref{star} to define the deformed exterior product by the formula 
\be 
\omega \wg_\star \omega' = \wg \, \big(F^{-1} \tri (\omega \ot \omega'\, ) \big) = \big( F_{(1)}^{-1}\tri \omega \big) \wg \big(F_{(2)}^{-1} \tri \omega'\, \big) \ . \label{ws}
\ee
The exterior derivative $\dd$ is still a derivation for the deformed exterior product, and this defines the {nonassociative differential calculus} on $R$-space. 

In this contribution we wish to apply this machinery to formulate and study field theory on nonassociative $R$-space. In order to set up a Lagrangian formalism, we need a suitable definition of integration on the subalgebra $\mc S({\cal M}) \subset C^\infty({\cal M})$ of Schwartz functions on ${\cal M}$ with the deformed product $\star$. For this, we observe that the nonassociative star product \eqref{sp} satisfies $ f \star g = f \, g + {(\mbox{total derivatives})}$, so that the usual integral on ${\cal M}$ satisfies the {2-cyclicity condition} 
\be 
\int_{{\cal M}}\, {\dd^{2d}x \ f \star g} =\int_{{\cal M}}\, {\dd^{2d}x \ g \star f} = \int_{{\cal M}}\, {\dd^{2d}x \  f\, g} \label{2cyclicity}
\ee
for all $f,g \in \mc S ({\cal M})$, i.e. it exhibits the correct classical behaviour and can thus be used on the deformed algebra of functions. Similarly, the star product of three fields satisfies the {3-cyclicity condition}
\be
\int_{{\cal M}}\, {\dd^{2d}x \ f \star (g \star h)} = \int_{{\cal
    M}}\, {\dd^{2d}x \ (f \star g) \star h} \ ,
\label{3cyclicity}
\ee
in harmony with the expectations that on-shell conformal field theory
correlation functions should see no traces of nonassociativity~\cite{Blumenhagen:2014sba,Mylonas:2014aga}.
Analogous graded cyclicity conditions also hold for differential forms
on ${\cal M}$ of arbitrary degree~\cite{Mylonas:2013jha}. However,
since a bracketing has to be specified for the nonassociative star
product of $n>3$ fields, the 2-cyclicity condition does \emph{not}
imply the usual cyclicity property of associative star products. But
with the use of 3-cyclicity, together with the pentagon relations for
the associator $\phi_F$, a classification of the equivalent integrated
bracketed expressions can be carried out, and one finds that the different ways of bracketing an integrated $n$-fold product of fields are organised into $C_{n-2}$ classes, one for each different bracketing where exactly one field sits outside of the brackets; here $C_n = \frac{(2n)!}{n!\, (n+1)!}$ are the Catalan numbers. These classes are related to one another by cyclic permutations, and every class of an integrated $n$-fold product of fields is mapped to another one by a cyclic permutation of the fields. 

These observations become particularly important when we study nonassociative field theories on constant $R$-flux backgrounds; here we focus on scalar field theories. The action functional of a free Euclidean scalar field $\varphi \in \mc S({\cal M})$  of mass $m$ on $R$-space is given by
\be 
S_0 = \frac12\, \int_{{\cal M}} \, \dd^{2d}x \ \big( \p _I \varphi \star \p^I \varphi + m^2 \, \varphi \star \varphi \big) \ . \label{S0}
\ee
Using 2-cyclicity the star products go away and $S_0$ becomes the standard free scalar field theory action on ${\cal M}$. This implies that the bare propagator is not affected by the nonassociative deformation, and thus similarly to the usual noncommutative field theories we have to consider interactions in order to probe nonassociative effects (see e.g.~\cite{Szabo:2001kg}). 
When introducing interactions the nonassociativity of the star product forces us to include all different bracketings of the product of $n$ fields in the action. However, due to the classification above we keep only one representative term from each class. For a single scalar field it follows from \eqref{2cyclicity} and \eqref{3cyclicity} that some of the classes are equivalent, while for $n=3,4,5$ all different bracketings of the interaction term can be shown to be equal. This means that the $\varphi^3$, $\varphi^4$ and $\varphi^5$ theories are associative at tree level and thus one should study loop corrections in order to detect nonassociativity. The first encounter of nonassociativity at tree level is for the $\varphi^6$ theory where four inequivalent interactions appear~\cite{Herbst:2002fk}.

Nonassociative interactions turn out to be rather tricky to deal with as they feature novelties that are not present in their associative counterparts. Since the geometrization of $R$-space is a phase space, Fourier modes for both configuration and momentum spaces have to be considered in the field expansions; we denote by the $2d$-vector $k_I=(\ka_i, l^i)$ the Fourier momenta corresponding to the phase space coordinates $x^I=(x^i, p_i)$. It is then straightforward to use the standard Fourier transform on each of the $C_{n-2}$ interaction terms and calculate the vertex phase factor. For example, in the case of the equivalence class given by the interaction term 
\be
S_{\trm{int}}^{(n)}= \frac {g}{n!} \, \int_{\cal M}\, \dd^{2d} x \ \big[ \cdots \big[\big( (\varphi \star \varphi ) \star \varphi\big) \star \varphi \big] \star\cdots \star \varphi\big] \ , \label{Sintn}
\ee
we can use 2-cyclicity \eqref{2cyclicity}, the star product \eqref{sp} and the Baker--Campbell--Hausdorff formula to arrive at the phase factor
\be
V\big(x,k^{(a)}\big)= \exp\bigg[ \ii \sum_{a=1}^{n}\, {k^{(a)}_I\, x^I} - \frac{\ii\hbar}{2} \, \sum_{1\leq a<b\leq n}\, {\Big( k_I^{(a)}\, k_J^{(b)}  \, \Theta^{IJ} + \frac{\hbar}{2} \, k_I^{(a)} \, k_J^{(b)} \, k_L^{(c)} \,  R^{IJL} \Big)}\bigg] \ ,  \label{Vn}
\ee
where $\big(\Theta^{IJ}\big)= \big({}^{R^{ijk}\, p_k}_{-\delta_i{}^j}
\ {}^{\delta^i{}_j}_0 \big)$ is the phase space deformation parameter
matrix and $\big( R^{IJK} \big) =\big({}^{R^{ijk}}_{ \ \ 0} \ {}^0_0\big)$ is the
Jacobiator matrix. The first two terms of \eqref{Vn} are the familiar
phase factor modification to the Feynman rules that appear in
associative noncommutative $\varphi^n$ theories, while the last term
is due to the nonassociative deformation.

An interesting feature of \eqref{Vn} is that it induces violation of momentum conservation at the vertex. This is due to the $x^I$ dependence of the deformation parameter and it is completely analogous to what occurs in the usual noncommutative field theories with spacetime varying noncommutativity parameter~\cite{Robbins:2003ry}. For this, we Fourier transform the fields in \eqref{Sintn} and then perform the integral over $\cal M$ to obtain the momentum relations
\be
\sum_{a=1}^{n}\, {\ka_i^{(a)}}=0 \qquad \trm{and} \qquad \sum_{a=1}^{n}\, {l^{(a) \, k}} = \frac{\hbar}{2} \ \sum_{1\leq a<b\leq n}\, { \ka_i^{(a)}\, \ka_j^{(b)} \, R^{ijk}} \ , \label{momentum}
\ee
where the first equation is the usual momentum conservation on configuration space while the second equation exhibits violation of momentum conservation along the noncommutative momentum space directions. 

An even more intriguing feature is the fact that the phase factor \eqref{Vn} is not invariant under cyclic permutations of the Fourier momenta. This is not obvious at first sight, but taking a cyclic permutation of the indices and using \eqref{momentum} reveals that the $R$-dependent block of the deformation parameter combines with $l$-momentum to violate cyclicity of the interaction. This novel feature is particular to nonassociative interactions and has drastic effects on the classification of Feynman diagrams. 

For this, regard a connected Feynman diagram as an abstract connected graph $G$ realised by its embedding in an orientable surface of genus $h$. The vertices of the graph represent spacetime points and the edges represent propagators. For commutative scalar field theories all vertices are indistinguishable and so are all edges, therefore edge crossings can always be avoided and all possible Feynman diagrams are given by planar graph embeddings, i.e. they can be drawn on a surface of genus $h=0$. Counting all different ways that vertices and edges can be put together on a plane provides the number of topologically equivalent diagrams which enters as the symmetry factor of the graph in the perturbation series. 

For the usual noncommutative deformations the interchange of two edges is no longer allowed and thus planarity cannot always be accomplished. Some symmetry remains though as the phase factor of the interaction is invariant under cyclic permutations of the edges. The Feynman diagrams are then classified by the minimal genus of the surfaces in which they are embedded; in fact, it is the cyclicity of the phase factor which guarantees that every way of embedding a graph $G$ into an orientable surface $\Sigma_h$ of genus $h$ is equivalent.

For nonassociative scalar field theories the vertex interaction is no
longer cyclic, which suggests that the different ways of embedding a
graph into  $\Sigma_h$ are no longer equivalent. This is supported by
the {rotational embedding scheme} proposed by
Edmonds~\cite{Edmonds:1960} and discussed in detail by
Youngs~\cite{Youngs:1963}. According to this theorem the 2-cell embedding of
a graph $G$ into $\Sigma_h$ is uniquely determined by a collection of
cyclic permutations $(\s_1, \dots,\s_v)$, where $\s_i$ is a
counterclockwise cyclic permutation of the edges connected to the
$i$-th vertex of $G$. The genus of the embedding surface $\Sigma_h$ is
given by $h = 1-\frac12 \, \big( \|V(G)\|-\|E(G)\|+\|O(\s_i)\| \big)
$, where $V(G)$ is the
set of vertices of $G$, $E(G)$ is the set of edges and $O(\s_i) ,\,
i=1,\dots,v$ is the set of orbits for the given $v$-tuple of
cyclic permutations. This is of course the
familiar Euler formula $2-2h = v-n+f $ for the embedding of a Feynman
diagram with $v$ vertices, $n$ edges and $f$ faces into an orientable surface of
genus $h$. The minimal embedding of the graph is given by the rotation
systems which provide the maximum number of orbits. Clearly if the
interaction vertex is not cyclic, then all different rotational
schemes that embed a graph into $\Sigma_h$ are potentially
inequivalent. However, cyclic permutations are also maps between
classes of integrated $n$-fold products of fields, i.e. between the edges connected to a vertex. We can use this residual symmetry to give a classification for the Feynman diagrams of nonassociative field theories: It is the standard noncommutative classification into planar and non-planar graphs where every Feynman diagram of genus $h$ with $v$ vertices and $n$ legs is subdivided into $v\, C_{n-2}$ inequivalent diagrams.

As an example, let us set $n=4$ in \eqref{Sintn} and \eqref{Vn} and
study nonassociative $\varphi^4$ theory on flat $R$-space. There are
$C_2=2$ classes of interaction terms which in this case are equal, but
their phase factors are related by $ V_\s (x,k^{\s (a)})=\exp\big(
\mbox{$\frac {\ii\hbar}{2}$} \, k_I^{(1)}\, k_J^{(2)}\, k_K^{(3)}  \,
R^{IJK} \big) \, V(x,k^{(a)}) $, where $\s\in S_4$ is a cyclic
permutation of the four indices. It follows that each phase factor is
invariant under the composition of two cyclic permutations, so that each diagram is subdivided into $2v$ classes. This suggests that there are two different 2-cell embeddings for each vertex, i.e. the off-shell field theory has two types of vertices, and thus planar and non-planar diagrams will accordingly split into subclasses determined by which permutation is used to embed each vertex.

Let us now calculate the one-loop corrections to the two-point
function in this model. In this case the contributions from the two subclasses are equal as the $R$-dependent term vanishes due to antisymmetry of $R^{IJK}$. 
The planar part is given by 
\be 
\Gamma^{(2)}_{\rm pl}= \frac {g}{3(2\pi)^{4d}} \, \int_{\cal M}\, {\dd^{2d}x} \ \int_{\cal M^*} \, \frac{\dd^{2d} \lambda}{\lambda^2 +m^2} \, V\big(x,k^{(2)}, k^{(3)} \big) \ , \label{planar}
\ee
where $k^{(2)}$, $k^{(3)}$ are the external momenta, $\lambda$ is the loop momentum and the phase factor is given by \eqref{Vn} with $n=4$. We now integrate over $x$ and use Schwinger parameters to turn the amplitude into a Gaussian integral, which yields the familiar commutative result 
\be 
\Gamma^{(2)}_{\rm pl}= \frac {g}{3 (2\pi)^d} \,
\delta^{(2d)}\big(k^{(2)}-k^{(3)}\big) \, \Big(\, \frac{m\, \Lambda}{2} \, \Big)^{d-1} \, K_{d-1}\big(\mbox{$\frac{2m}{\Lambda}$} \big) \ , \label{pl} 
\ee
where $K_n$ is the modified Bessel function of the second kind of
order $n$ and $\Lambda$ is an ultraviolet cutoff which has been
introduced to regularize the Schwinger integral.

For the non-planar diagram it is convenient to postpone the integral
over $\cal M$ as it gives a delta-function constraint which is singular and cannot be used to evaluate the remaining integrals. Instead we introduce a Schwinger parameter and integrate over the loop momentum which yields
\be
\begin{split}
\Gamma^{(2)}_{\rm np}=& \ \frac{g \, \pi^d}{6(2\pi)^{4d}} \,
\int_0^\infty\, \frac{\dd^d s}{s^d} \ \exp \Big( -s \, m^2 -
\frac{\hbar^2 \, \big(k^{(2)}\big)^2}{4 s}\, \Big) \ \int_M\, \dd^{d}x \ \exp \big( \ii (\ka^{(2)}- \ka^{(4)} )_i \, x^i\ \big)   \\
& \ \times \ \int_{(\real^d)^*}\, \dd^d p \ \exp \Big(- \frac{\hbar^2}{4 s}\, A^{mn}\,p_m \, p_n + \ii \big(l^{(2)}- l^{(4)} \big)^i \, p_i \Big) \ , \label{midway}
\end{split}
\ee
where $A=(A^{mn}):= \sum_{k=1}^d \, \ka_i \, \ka_j \, R^{ikm} \,
R^{jkn}$ is a singular symmetric $d\times d$ matrix. In this
expression the integral over configuration space $M$ yields a
delta-function while the momentum space integral is Gaussian. Since
$A$ {is} singular, it has $p \geq1$ zero eigenvalues and $d-p$
non-zero real eigenvalues $\rho_1,\ldots, \rho_{d-p}$. Thus only $d-p$
independent momentum space integrals are Gaussian while the rest yield delta-functions. With these manipulations we can integrate over $x$ and $p$ in \eqref{midway}; we then introduce an ultraviolet cutoff $\Lambda$ and integrate over the Schwinger parameter $s$ to obtain
\begin{eqnarray}
\Gamma^{(2)}_{\rm np}&=& \frac{g}{6 \hbar^{d-p} \, 2^p \, (2\pi)^{\frac{3d-p}{2}} \, \sqrt{\rho_{1}
\cdots \rho_{d-p}} } \ \delta^{(d)} \big(\ka^{(2)} - \ka^{(4)}\big) \
\prod_{j=1}^{p}\, \delta^{(d)} \big((l^{(2)} - l^{(4)}) \cdot e_j \big)
\nonumber \\ && \ \qquad \times \ \Big(\, \frac{m_{\trm{eff}}\,
  \Lambda_{\trm{eff}}}{2} \, \Big)^{\frac{d-p}{2} -1} \, K_{\frac{d-
    p}{2}-1} \big(\mbox{$\frac{2m_{\trm{eff}}}{\Lambda_{\trm{eff}}}$} \big) \ ,
\label{nonplanar}\end{eqnarray}
where $e_1, \ldots, e_d$ are the eigenvectors of $A$, while the effective cutoff and mass are given by
\be
\frac{1}{\Lambda^2_{\trm{eff}}}=\frac{1}{\Lambda^2} + \Big( \,
\frac{\hbar \, k^{(2)}}{2} \, \Big)^2  \qquad \trm{and} \qquad
m^2_{\trm{eff}}=  m^2 +  \frac{4}{\hbar^2} \ \sum_{j=p+1}^d \,
\frac{\big(l^{(2)} -l^{(4)}\big) \cdot e_j}{\rho_{j-p}} \ .
\ee

It is not surprising that the one-loop contribution \eqref{nonplanar}
is qualitatively similar to the one calculated for ordinary
noncommutative scalar field theory with non-constant deformation
parameter; it even exhibits the usual UV/IR mixing
pathologies~\cite{Robbins:2003ry}. In general, for $n=3,4,5$ the
nonassociative $\varphi^n$ theories are on-shell associative, and
therefore the only nonassociative effects that they exhibit is in the
subdivision of their vertices into $C_{n-2}$ classes. Although this
was not observed in our example, it can be seen in higher loops and
even in the one-loop correction to the four-point function of the $\varphi^4$ theory.

We close by remarking that, while a formulation of quantum mechanics
in the phase space description of $R$-space is well-defined and
meaningful~\cite{Mylonas:2013jha}, the complete physical
interpretation of quantum
field theory on phase space is at the moment unclear. A statistical
approach is taken in~\cite{Amorim:2014uza} where it was
proposed that fields on phase space acquire a physical meaning via
an association to Wigner functions, but an interpretation related to deformation theory is still lacking. As the phase space formalism naturally arises in the geometrization of $R$-space, it is tempting to think of this model along the lines of double field theory; it is then interesting to understand if there is some analog of the section condition reducing the phase space field theory to field theory on the original configuration manifold $M$.\\

This work is supported in part by the Greek National Scholarship
Foundation and the Consolidated Grant ST/J000310/1 from the UK Science and Technology Facilities Council.

\end{document}